\documentclass[10pt,final,twocolumn,journal]{IEEEtran}
\usepackage{filecontents,lipsum}
\usepackage[noadjust]{cite}
\usepackage[percent]{overpic}
\usepackage{stfloats}
\usepackage{ctable}
\usepackage[mathlines]{lineno}

\usepackage{cite}
\usepackage{graphicx}
\usepackage{caption}
\usepackage{subcaption}
\usepackage{authblk}
\usepackage{xcolor}
\usepackage{amsmath}
\usepackage{array}
\usepackage[normalem]{ulem}

\begin{document}
\title{Can Tunnel Transistors Scale Below 10nm?} 
\author{Hesameddin Ilatikhameneh, Gerhard Klimeck and Rajib Rahman
\thanks{This work was supported in part by the Center for Low Energy Systems Technology (LEAST), one of six centers of STARnet, a Semiconductor Research Corporation program sponsored by MARCO and DARPA.}
\thanks{The authors are with the Department of Electrical and Computer Engineering, Purdue University, USA, e-mail: hesam.ilati2@gmail.com.}
}
\maketitle
\setlength{\textfloatsep}{7pt plus 1.0pt minus 2.0pt}
\begin{abstract}
The main promise of tunnel FETs (TFETs) is to enable supply voltage ($V_{DD}$) scaling in conjunction with dimension scaling of transistors to reduce power consumption. However, reducing $V_{DD}$ and channel length ($L_{ch}$) typically deteriorates the ON- and OFF-state performance of TFETs, respectively. Accordingly, there is not yet any report of a high performance TFET with both low V$_{DD}$ ($\sim$0.2V) and small $L_{ch}$ ($\sim$6nm). In this work, it is shown that scaling TFETs in general requires \emph{scaling down} the bandgap $E_g$ and \emph{scaling up} the effective mass $m^*$ for high performance. Quantitatively, a channel material with an optimized bandgap ($E_g\sim1.2qV_{DD} [eV]$) and an engineered effective mass ($m*^{-1}\sim40 V_{DD}^{2.5} [m_0^{-1}]$) makes both $V_{DD}$ and $L_{ch}$ scaling  feasible with the scaling rule of $L_{ch}/V_{DD}=30~nm/V$ for $L_{ch}$ from 15nm to 6nm and corresponding $V_{DD}$ from 0.5V to 0.2V. 

\end{abstract}

\section{Introduction}
Although tunnel FETs (TFETs) were originally proposed for low power applications \cite{Appenzeller1, Appenzeller2, Ionescu}, the low ON-current (I$_{\rm ON}$) challenge in TFETs has concealed their scaling problem \cite{Ian, Ionescu2, Seabaugh}. The low I$_{\rm ON}$ challenge can be solved by increasing the electric field at the tunnel junction; e.g. by using dielectric engineering \cite{Hesam3}, atomistically thin channels \cite{Hesam1, Fiori, Hesam2, Fiori2}, or internal polarization \cite{Wenjun}. However, the scaling challenge is more tricky since the tunneling currents I$_{\rm ON}$ and I$_{\rm OFF}$ depend on the same device parameters. Hence an attempt to decrease I$_{\rm OFF}$ would reduce I$_{\rm ON}$ and vice versa. In contrast, I$_{\rm ON}$ and I$_{\rm OFF}$ in MOSFETs are more independent of each other and a channel material with a large bandgap (or optimized effective mass) can be used for sub-12nm channels to suppress the direct source-to-drain tunneling \cite{Sub12, lake}. 

\begin{figure}[!t]
        \centering       
        \begin{subfigure}[b]{0.27\textwidth}
               \includegraphics[width=\textwidth]{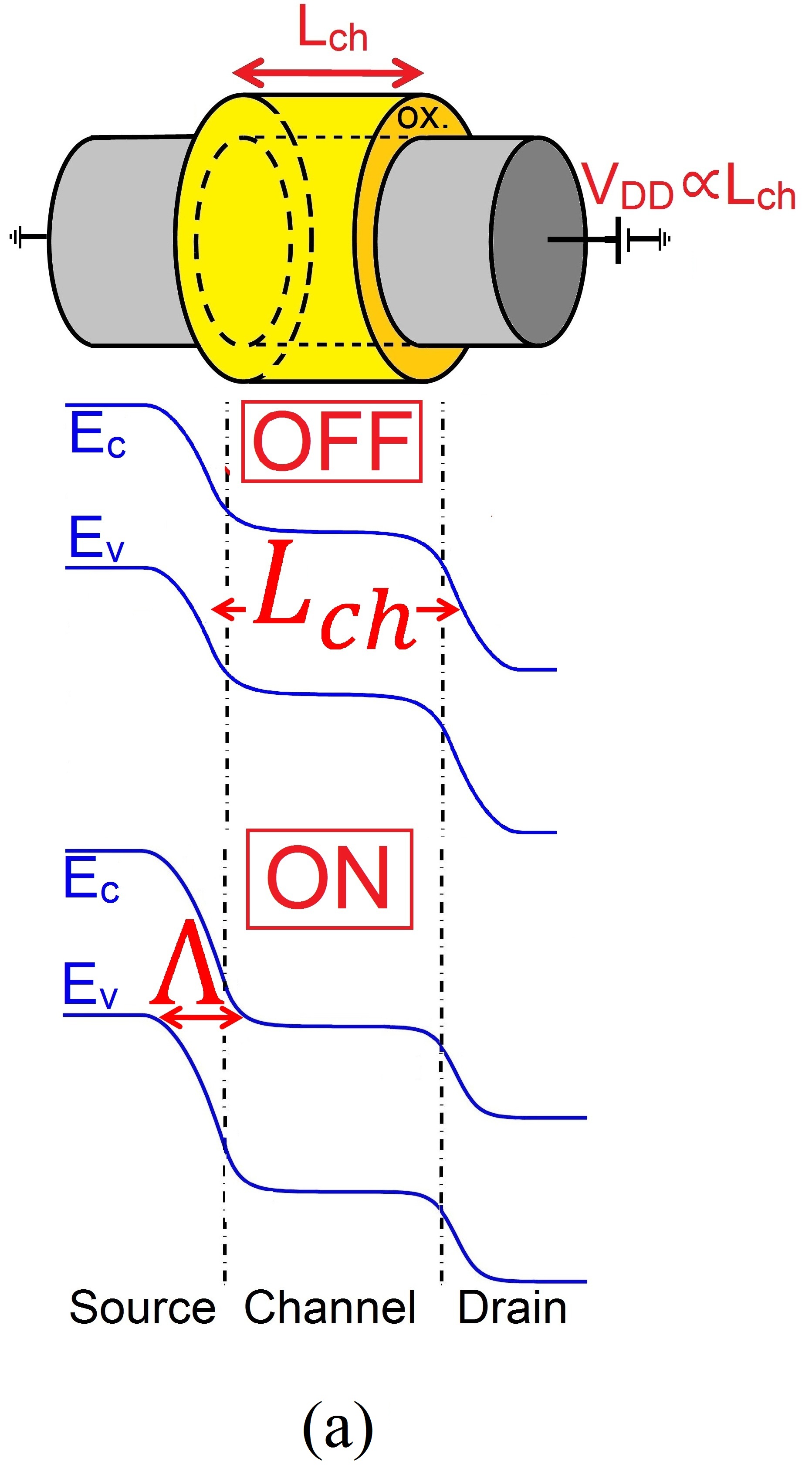}
                \label{fig:struct1}
        \end{subfigure}%
        \begin{subfigure}[b]{0.24\textwidth}
               \includegraphics[width=\textwidth]{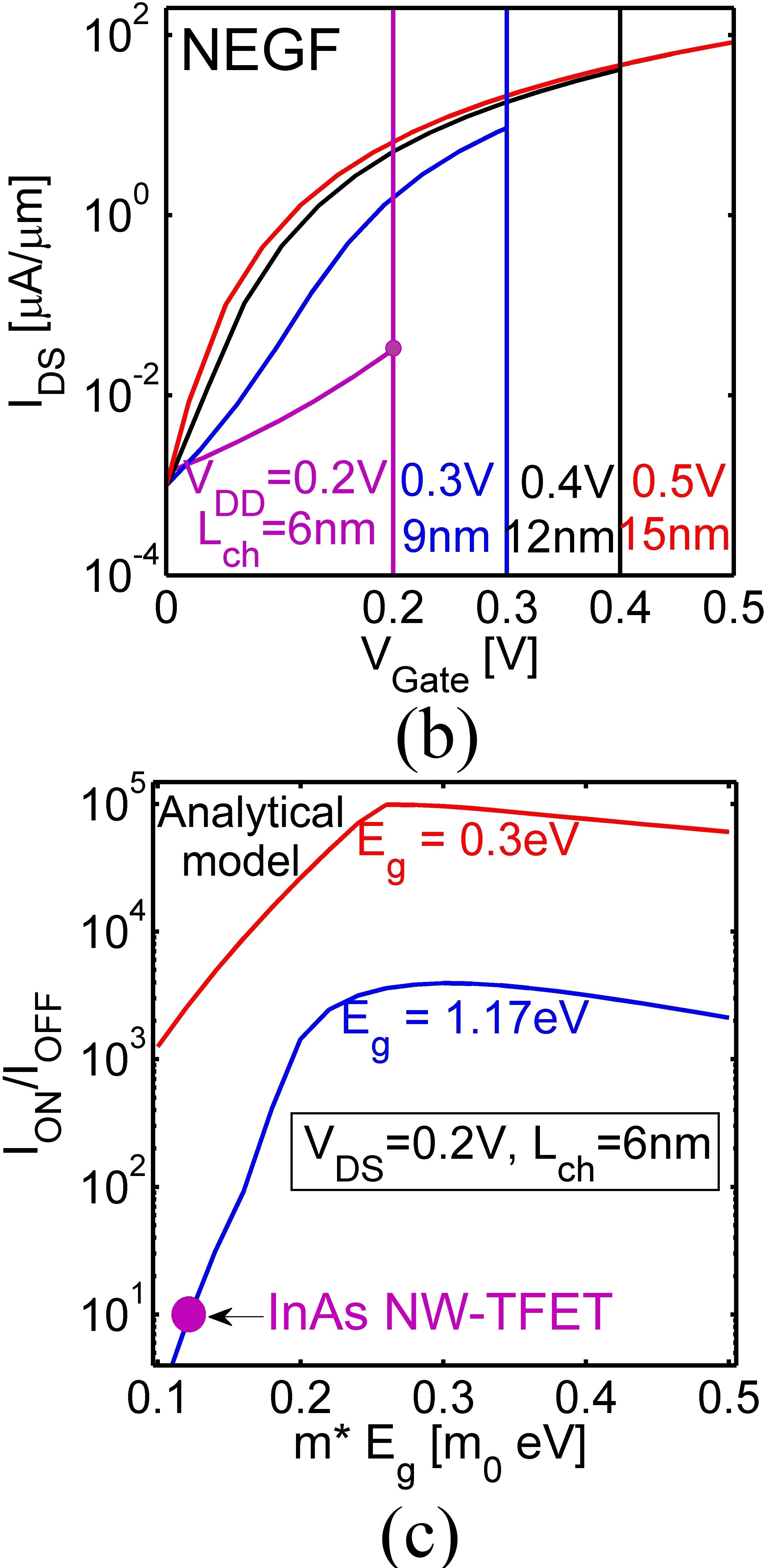}
                \label{fig:Same_EoT}
        \end{subfigure}%
        \vspace{-1.5\baselineskip}
        
        \caption{a) Structure of a NW-TFET with an ideal voltage scaling \cite{Ideal}. The band diagrams depict the ON and OFF states. b) The transfer characteristics of InAs nanowire TFET with the scaling rule of $L_{ch}/V_{DD}=30~ nm/V$ for channel lengths from 15nm to 6nm which translates into corresponding $V_{DD}$ from 0.5V to 0.2V (from NEGF). A 3nm thick HfO$_2$ with $\epsilon_{ox}$=25 has been used as the gate oxide. c) Impact of $m^*$ and $E_g$ variations on I$_{\rm ON}$/I$_{\rm OFF}$ for a TFET with $L_{ch}$=6nm and $V_{DD}$=0.2V predicted from an analytical model (see Sec. \ref{sec:sim}).}\label{fig:Fig1}
\end{figure}

Fig. \ref{fig:Fig1}a shows the device structure of an InAs nanowire (NW) TFET with a diameter of 3.4nm. The transfer characteristics of the device simulated by the NEMO5 tool \cite{nemo5_1, nemo5_2, nemo5_3} are shown in Fig. \ref{fig:Fig1}b with I$_{\rm OFF}$ fixed at 1$nA/\mu m$. In the simulations, we scale $V_{DD}$ down with the channel length ($L_{ch}$). The results indicate that the InAs NW-TFET exhibits a promising performance with long channel lengths (i.e. $L_{ch}>$ 9nm), however it completely fails to switch from OFF- to ON-state for the case of $L_{ch}$=6nm and $V_{DD}$=0.2V (i.e. I$_{\rm ON}$/I$_{\rm OFF}\approx 10 \ll 10^4$). 

Roughly, the transmission in the ON-state ($T_{\rm ON}$) and OFF-state ($T_{\rm OFF}$) of TFETs depends on \cite{Analytic2, Kane}:
\begin{equation}
\label{eq:tun1}
log(T_{\rm ON}) \propto \Lambda \sqrt{m^*_r E_g}
\end{equation}
\begin{equation}
\label{eq:tun2}
log(T_{\rm OFF}) \propto L_{ch} \sqrt{m^*_r E_g}
\end{equation}
\noindent where $\Lambda$ and $L_{ch}$ are the tunneling distances in the ON- and OFF-state (Fig. \ref{fig:Fig1}a) respectively. $m^*_r$ and $E_g$ are the reduced effective mass and the bandgap of the channel material. 

The scaling of the channel below 10nm brings $L_{ch}$ close to $\Lambda$ which reduces I$_{\rm ON}$/I$_{\rm OFF}$ significantly. One apparent solution can be a heterostructure channel where the term $m^*_r E_g$ is different in (\ref{eq:tun1}) and (\ref{eq:tun2}) due to different materials used in those regions \cite{Wenjun, Het1}. However, it has been shown that the presence of band discontinuity and interface states in heterostructures can deteriorate the OFF-state performance of TFETs \cite{Het2, Het3}. Hence, in this work the homojunction TFETs have been considered as a more practical steep subthreshold swing ($SS$) device.

On top of the length scaling problem which increases I$_{\rm OFF}$ significantly, the voltage scaling reduces I$_{\rm ON}$. The maximum tunneling window in TFETs approximately equals $qV_{DD}$. Thus a short channel TFET with a small $V_{DD}$ is expected to have a small I$_{\rm ON}$/I$_{\rm OFF}$. 

In this work, it is shown that by using a channel material with optimized $m^*$ and $E_g$, it is still feasible to obtain an acceptable I$_{\rm ON}$/I$_{\rm OFF}$ for ultra-scaled TFETs (i.e. I$_{\rm ON}$/I$_{\rm OFF} > 10^5$ for $L_{ch}$=6nm and $V_{DD}$=0.2V). The solution to the scaling problem of TFET is to \emph{scale down} $E_g$ of channel material to the smallest possible value to achieve a high I$_{\rm ON}$. Of course $E_g$ cannot be smaller than $qV_{DD}$, otherwise the channel cannot cover and block the tunneling energy window in the OFF-state. On the other hand, $m^*$ should \emph{scale up} with scaling down the dimensions to decrease I$_{\rm OFF}$. Fig. \ref{fig:Fig1}c shows that the performance of 6nm long gate-all-around TFET can be improved more than 4 orders of magnitude by \emph{scaling down} $E_g$ and \emph{scaling up} $m^*$. The favorable design space for $m^*$ and $E_g$ is discussed in Sec. \ref{sec:opt}. 
  
\section{Simulation details}
\label{sec:sim}
\begin{figure}[!t]
        \centering    
        \begin{subfigure}[b]{0.35\textwidth}
               \includegraphics[width=\textwidth]{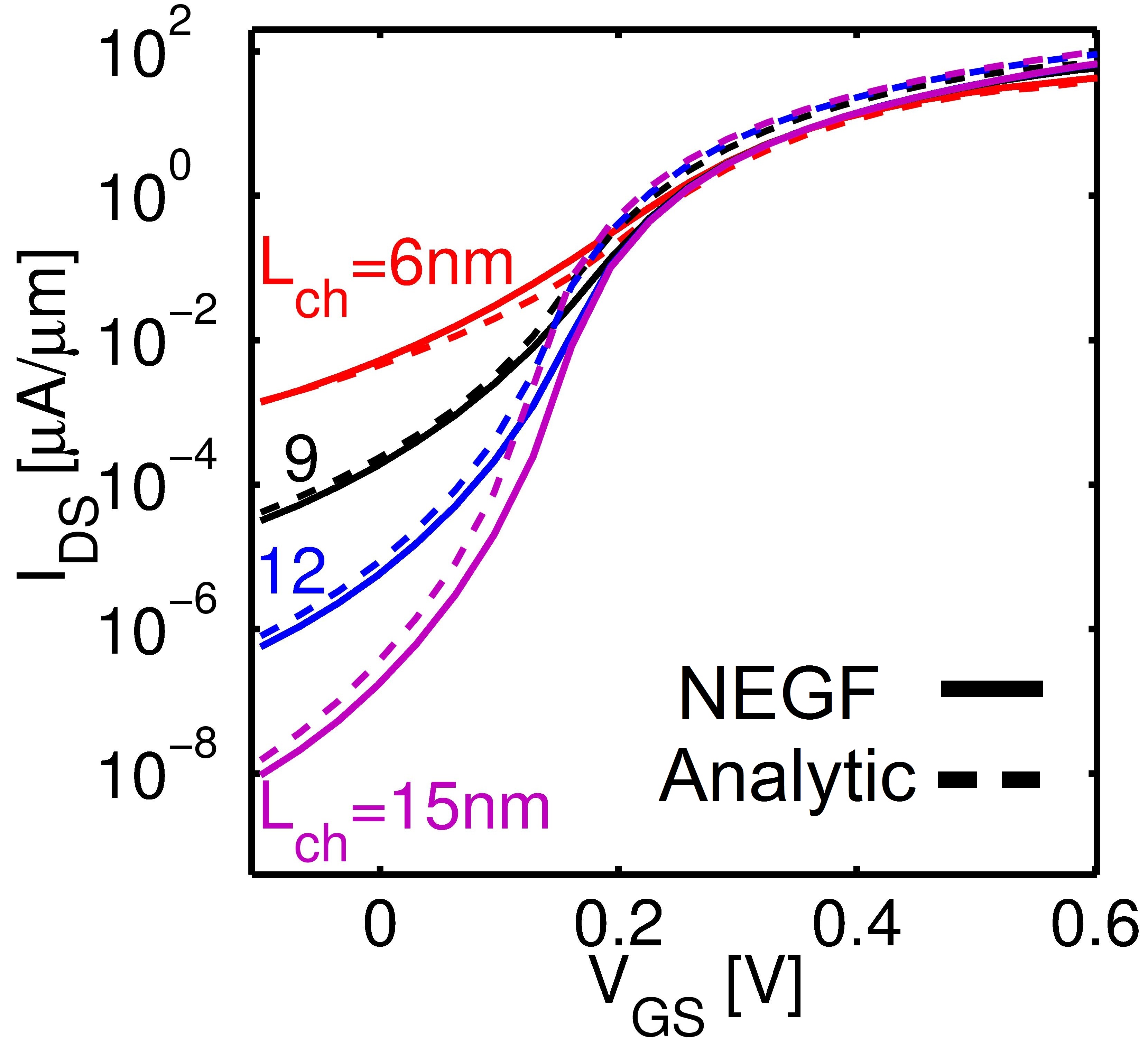}
               \vspace{-1.3\baselineskip}
                \label{fig:Same_EoT}
        \end{subfigure}%
        \vspace{-0.3\baselineskip}
        \caption{The comparison between $I_{D}$-$V_{G}$ of the InAs NW-TFET obtained from analytical model (dashed lines) and NEGF simulations (solid lines) with the same parameters as Fig. \ref{fig:Fig1}. \label{fig:Fig2}}
\end{figure}

The self-consistent 3D Poisson-NEGF (Non-Equilibrium Green's Function) method is used in the NEMO5 software for the simulation of InAs TFETs \cite{nemo5_1, nemo5_2, nemo5_3}. The InAs channel material is described by a 10 band nearest neighbor tight-binding model \cite{Boykin}. To find the impact of $m^*$ and $E_g$ on the performance of TFETs, a model is needed where $m^*$ and $E_g$ can be set as free input parameters, in contrast to the atomistic approach where $m^*$ and $E_g$ are the output of the simulation through material composition and geometry induced confinement effects. To reduce the number of free parameters, it is assumed that the electron and hole effective masses are equal ($m^*_e$=$m^*_h$=$m^*$). Recently, an analytical model was developed which produces results in excellent agreement with NEGF simulations \cite{Analytic1}. To show the validity of this analytical model for ultra-scaled TFETs, the simulation results of scaled InAs TFETs with the scaling rule of $L_{ch}/V_{DD}=30~ nm/V$ obtained from the analytical model are benchmarked against the NEGF results first. Fig. \ref{fig:Fig2} compares the results of analytical model and NEGF simulations. Notice that in Fig. \ref{fig:Fig2}, the OFF-state is not fixed unlike Fig. \ref{fig:Fig1}b. The accuracy and speed of the analytical model and tuneability of $m^*$ and $E_g$ makes this model an ideal tool for optimizing the TFET design.

\section{Simulation results}
\begin{figure}[!t]
        \centering    
        \begin{subfigure}[b]{0.25\textwidth}
               \includegraphics[width=\textwidth]{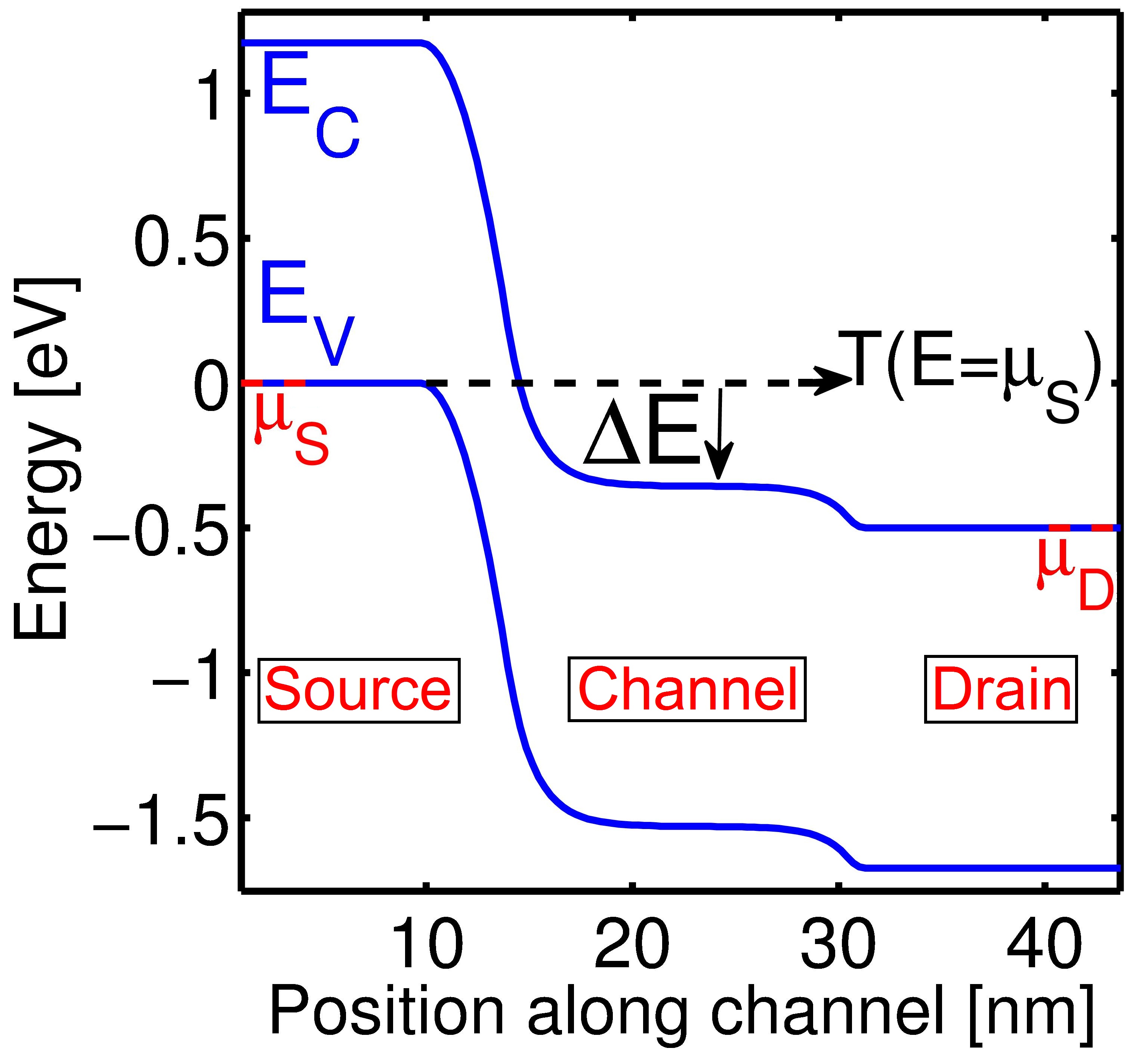}
               \vspace{-1.5\baselineskip}
                \caption{}
                \label{fig:Spacing}
        \end{subfigure}%
        \begin{subfigure}[b]{0.25\textwidth}
               \includegraphics[width=\textwidth]{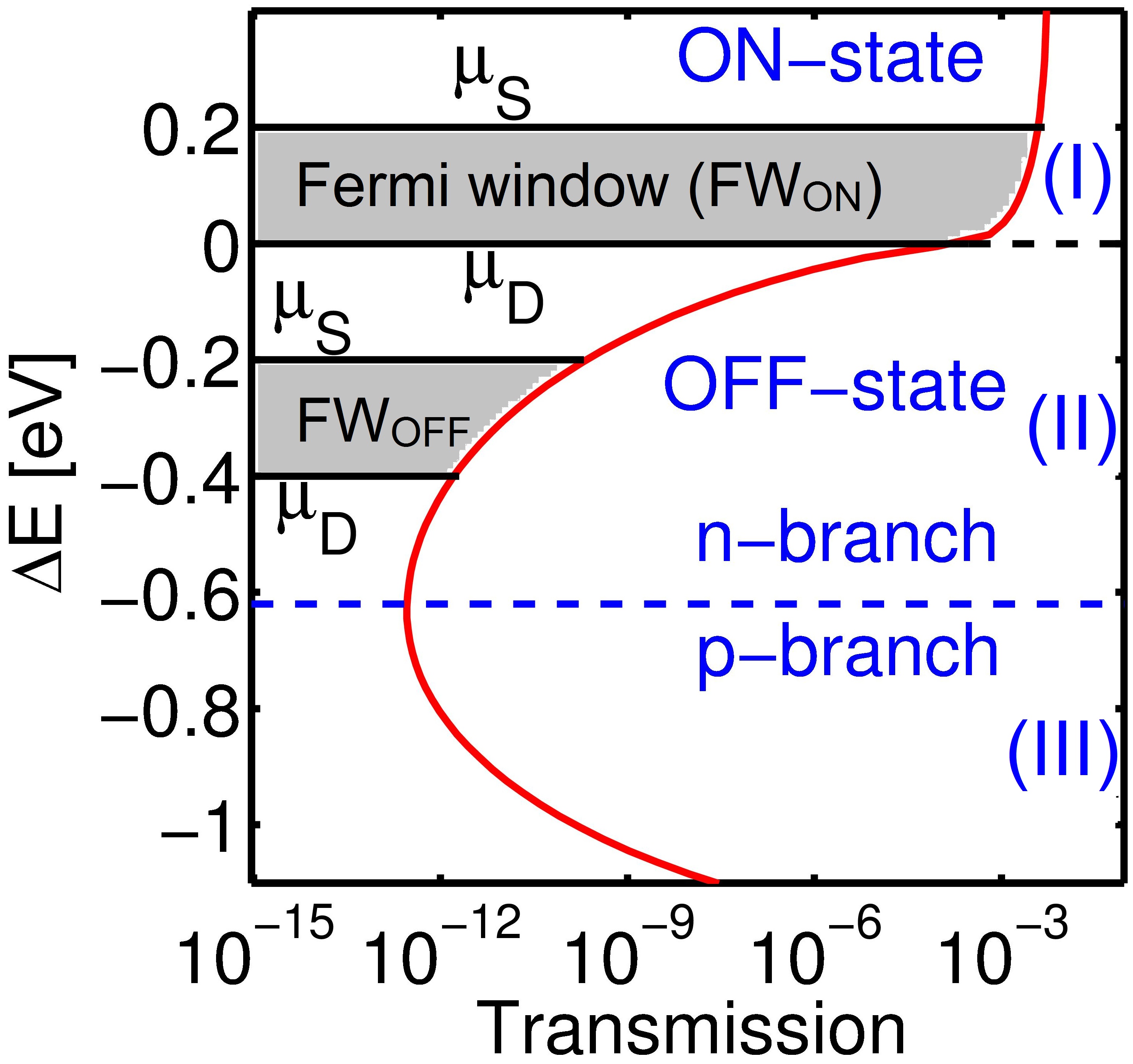} 
               \vspace{-1.4\baselineskip}
                \caption{}
                \label{fig:struct1}
        \end{subfigure}%
        \quad                    
        \begin{subfigure}[b]{0.25\textwidth}
               \includegraphics[width=\textwidth]{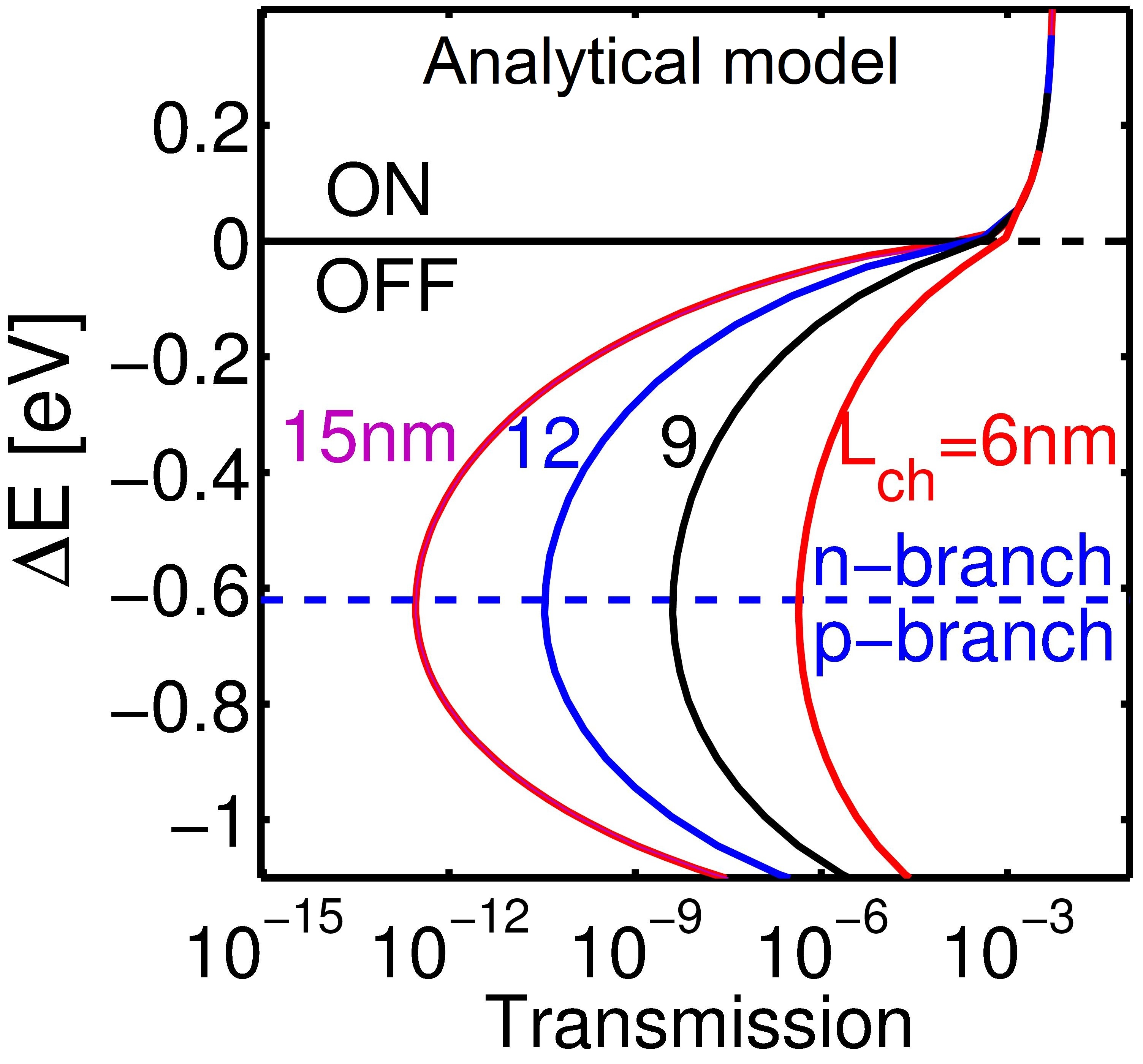}
               \vspace{-1.5\baselineskip}               
                \caption{}
                \label{fig:laplace}
        \end{subfigure}%
                \begin{subfigure}[b]{0.25\textwidth}
               \includegraphics[width=\textwidth]{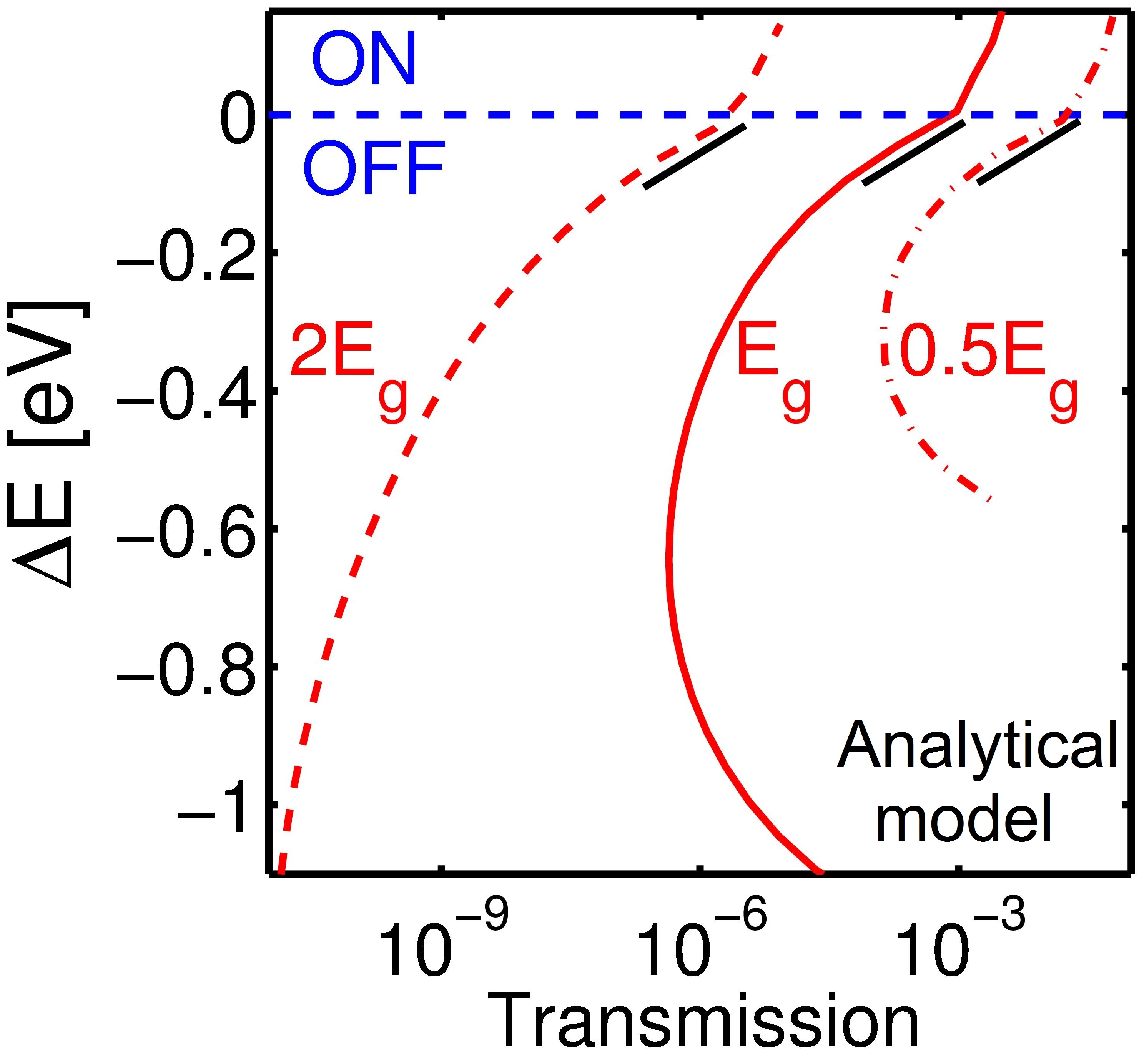}
               \vspace{-1.5\baselineskip}               
                \caption{}
                \label{fig:laplace}
        \end{subfigure}%
        \quad        
        \begin{subfigure}[b]{0.25\textwidth}
               \includegraphics[width=\textwidth]{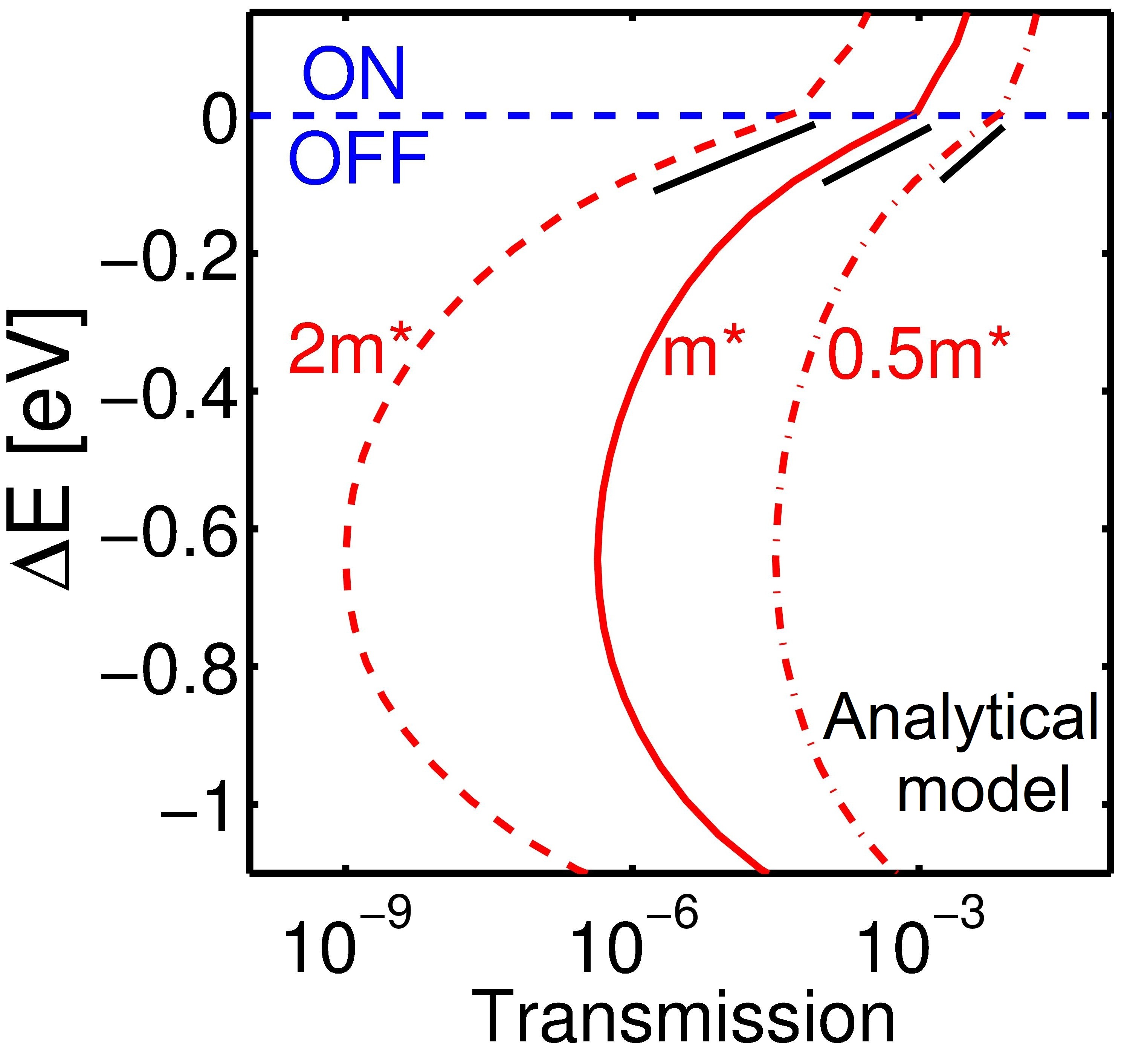}
               \vspace{-1.25\baselineskip}                              
                \caption{}
                \label{fig:st3}
        \end{subfigure}%
        \begin{subfigure}[b]{0.25\textwidth}
               \includegraphics[width=\textwidth]{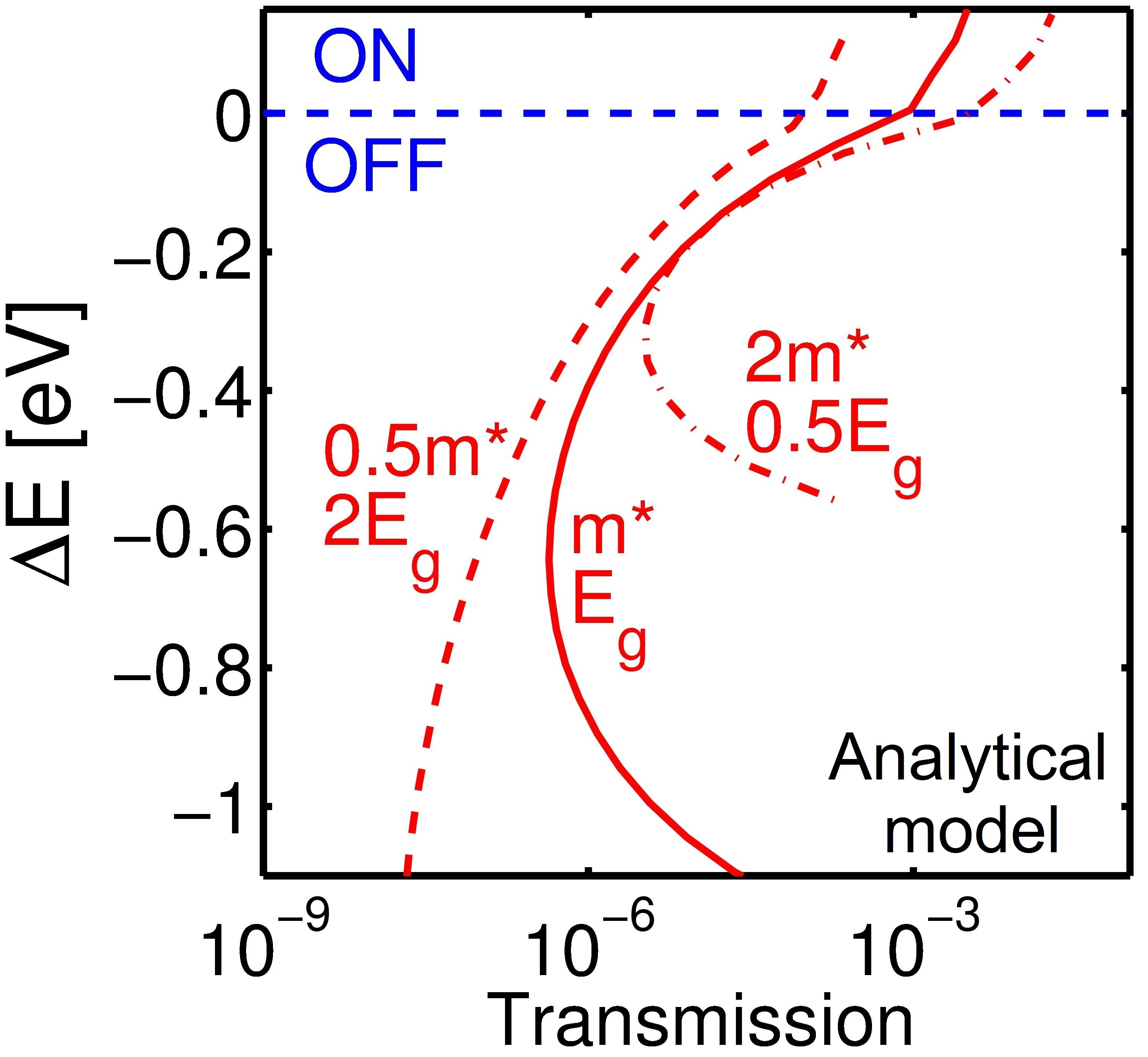}
               \vspace{-1.25\baselineskip}                              
                \caption{}
                \label{fig:st3}
        \end{subfigure}%
        \vspace{-0.3\baselineskip}
        \caption{a) Band diagram of a TFET showing $\Delta E$ parameter and b) the transmission profile $T(\Delta E)$. The impact of c) channel length $L_{ch}$, d) bandgap $E_g$, e) effective mass $m^*$, f) $m^*/E_g$ ratio on the transmission profile $T(\Delta E)$. All these results are obtained using the analytical model.}\label{fig:Fig3}
\end{figure}

To analyze different TFET designs, the tunneling transmission path at the top of the tunneling window ($E=\mu_S$) is indicated as a function of source-channel tunneling window ($\Delta E$) in Fig. \ref{fig:Fig3}a. Knowledge of the tunneling transmission probability as a function of $\Delta E$ (i.e. $T(\Delta E)$) provides information about the transfer characteristics \cite{Ionescu}. Fig. \ref{fig:Fig3}b shows an example of $T(\Delta E)$ with the corresponding TFET operational regimes (e.g. ON- and OFF-states, and n- and p- branches). Notice that $\Delta E \approx 0$ is the ON-OFF transition point. For a small drain-source voltage, the I-V can be calculated by integrating the $T(\Delta E)$ in the tunneling energy window (energies between $\mu_S$ and $\mu_D$). The tunneling transmission shows how far the TFET is from its ideal performance (i.e. $T=0$ and $T=1$ at OFF- and ON-state, respectively). Accordingly, I$_{\rm ON}$, I$_{\rm OFF}$, and $SS$ can be estimated from the maximum and minimum values of $T(\Delta E)$ and its slope at subthreshold region. The impact of $L_{ch}$ scaling on the transmission profile of InAs NW-TFET is shown in Fig. \ref{fig:Fig3}c. Reducing the channel length increases $T_{OFF}$ significantly while $T_{ON}$ remains intact which was expected from equations (\ref{eq:tun1}) and (\ref{eq:tun2}).
 
Fig. \ref{fig:Fig3}d shows the effect of bandgap on $T(\Delta E)$; Obviously, a larger bandgap decreases both  $T_{OFF}$ and $T_{ON}$. Notice that changing $E_g$ does not improve the subthreshold slope of $T(\Delta E)$ (black lines in Fig. \ref{fig:Fig3}d). Increasing $E_g$ decreases $T_{OFF}$ more than $T_{ON}$ since the prefactor of $\sqrt{m^*_r E_g}$ is larger for $T_{OFF}$ (note that $L_{ch} > \Lambda$ in equations (\ref{eq:tun1}) and (\ref{eq:tun2})). On the other hand, to reach this lower $T_{OFF}$ a larger gate voltage change is needed for larger band gaps (i.e. $\Delta E_{OFF} \approx -E_g/2$). Thus, there is no noticeable improvement in $SS$ with larger $E_g$. On the other hand, increasing $m^*$ improves $SS$ as shown in Fig. \ref{fig:Fig3}e. Since a larger $m^*$ does not require a larger gate voltage change, contrary to a larger $E_g$. Fig. \ref{fig:Fig3}f compares TFETs with a constant $\sqrt{m^*_r E_g}$ but different $m^*_r/E_g$ ratios. Notice that not only $SS$ improves with increasing $m^*_r/E_g$ ratio, but also $T_{ON}$.  The reason for improved ON-state performance is that reducing $E_g$ decreases the depletion width at the source-channel interface and $\Lambda$ decreases in equation (\ref{eq:tun1}) \cite{Analytic2}.

\section{Channel material with optimized properties}
\label{sec:opt}

\begin{figure}[!b]
        \centering
        \begin{subfigure}[b]{0.25\textwidth}
               \includegraphics[width=\textwidth]{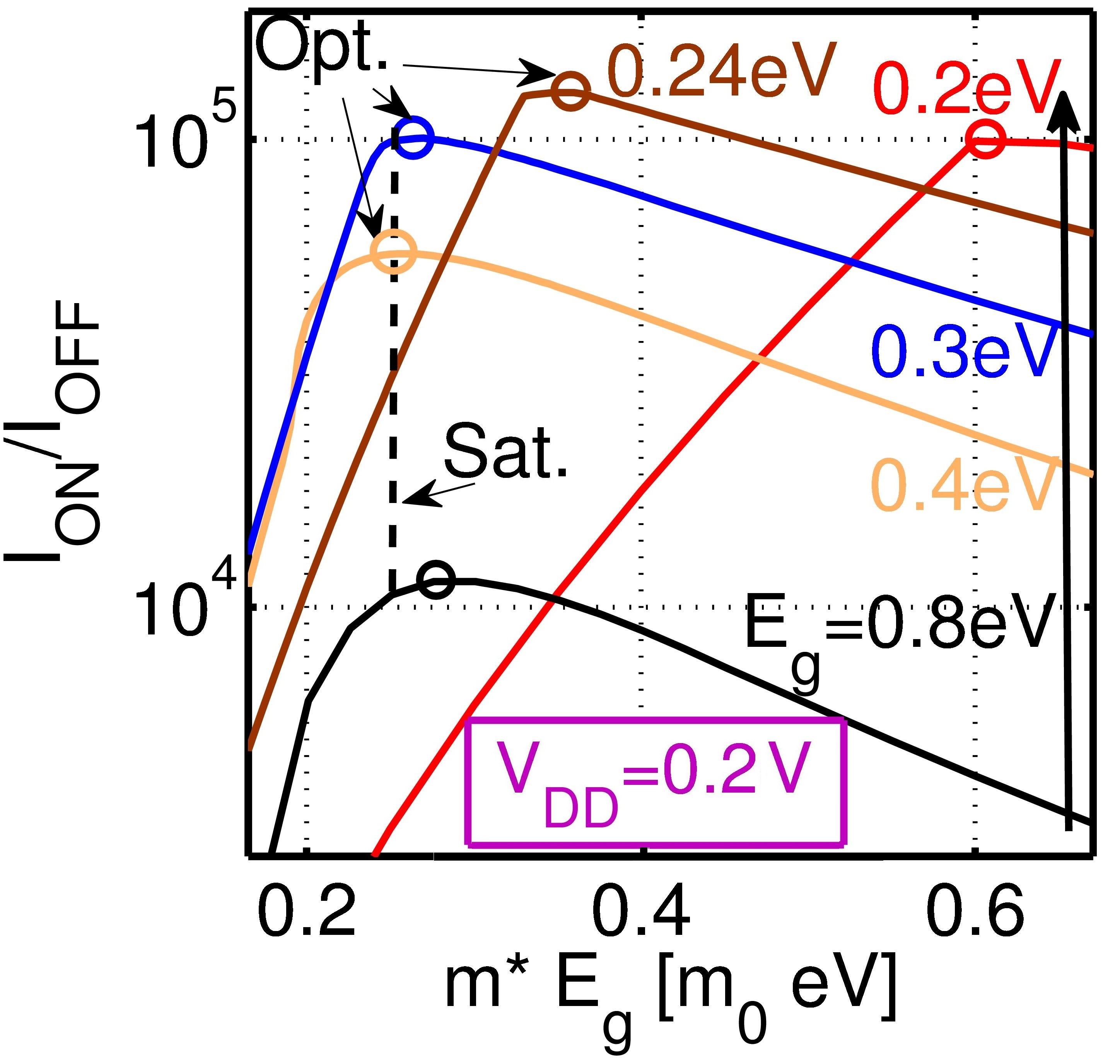}
               \vspace{-1.5\baselineskip}               
                \caption{}
                \label{fig:laplace}
        \end{subfigure}%
                \begin{subfigure}[b]{0.24\textwidth}
               \includegraphics[width=\textwidth]{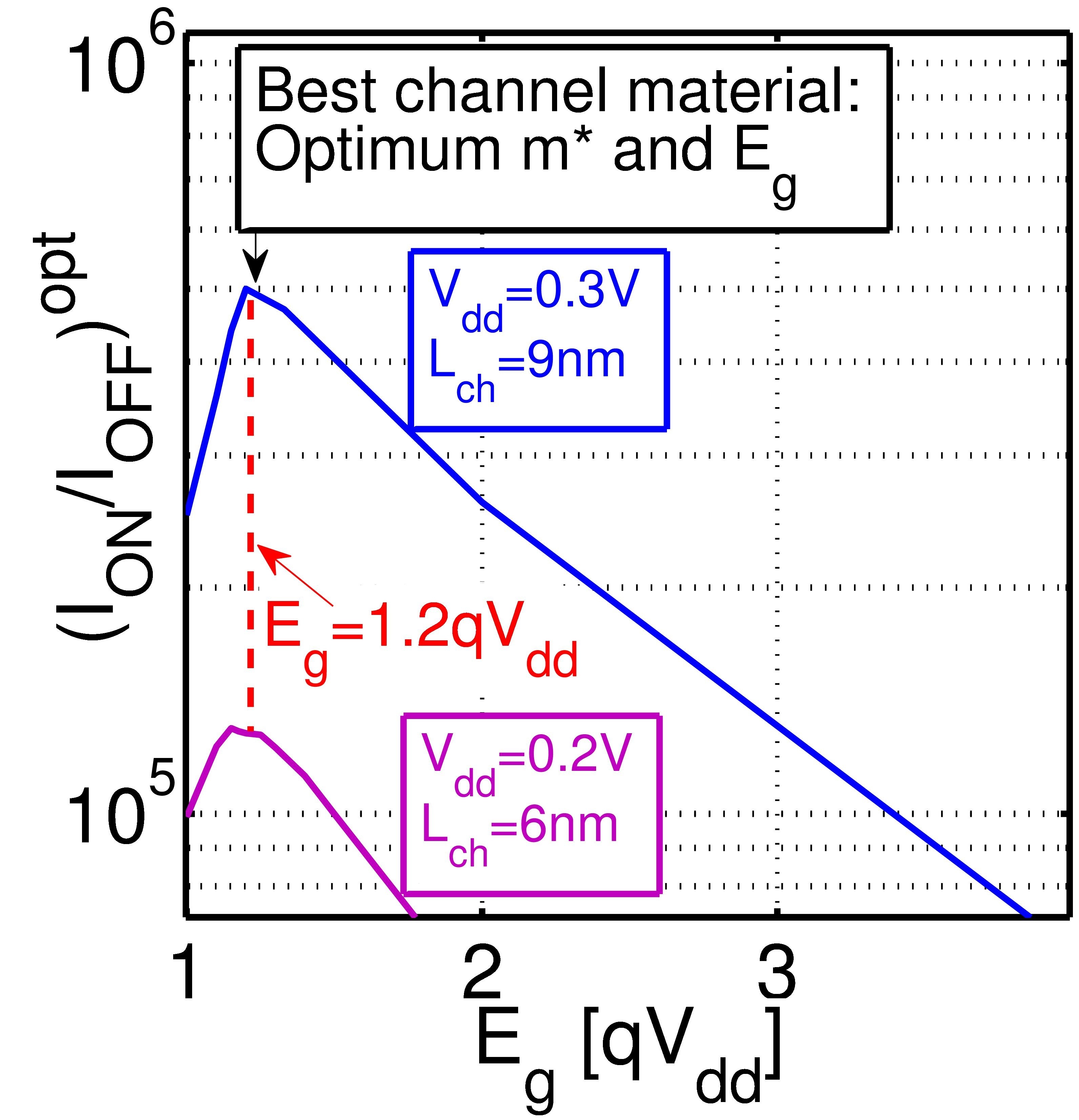} 
               \vspace{-1.5\baselineskip}               
                \caption{}
                \label{fig:laplace}
        \end{subfigure}%
        \vspace{-.5\baselineskip}       
        \caption{a) I$_{\rm ON}$/I$_{\rm OFF}$ ratio for a NW-TFET with $L_{ch}$=6nm and $V_{DD}$=0.2V for different $E_g$ and their optimized values $(I_{ON}/I_{OFF})^{opt}$ (circle symbols). b) $(I_{ON}/I_{OFF})^{opt}$ as a function of $E_g$ for $V_{DD}$=0.2V and 0.3V.}\label{fig:Fig4}	
\end{figure}

Fig. \ref{fig:Fig4}a shows the I$_{\rm ON}$/I$_{\rm OFF}$ ratio of NW-TFETs with $L_{ch}$=6nm and $V_{DD}$=0.2V and a channel material with different $m^*$ and $E_g$. To suppress the p-branch of TFETs, the drain doping level is chosen to be much smaller than source doping level ($N_S=20N_D=10^{20} cm^{-3}$) and a gate leakage of 1$nA/\mu m$ is assumed (I$_{\rm OFF} \geq 1nA/\mu m$) \cite{optK}. The maximum I$_{\rm ON}$/I$_{\rm OFF}$ ratio is obtained with an $E_g$ of about $1.2qV_{DD}$. Moreover, with increasing $E_g$, the optimum $m^*_{opt}$ reduces and for $E_g \geq 1.5qV_{DD}$ the product $m^*_{opt}E_g^{opt}$ (circle symbols) saturates (dashed line). Fig. \ref{fig:Fig4}b shows that TFETs with $E_g$ between $1.1qV_{DD}$ and $1.5qV_{DD}$ have acceptable I$_{\rm ON}$/I$_{\rm OFF}$ ratios according to ITRS requirements (I$_{\rm ON}$/I$_{\rm OFF}>10^5$).

\begin{figure}[!t]
        \centering
        \begin{subfigure}[b]{0.25\textwidth}
               \includegraphics[width=\textwidth]{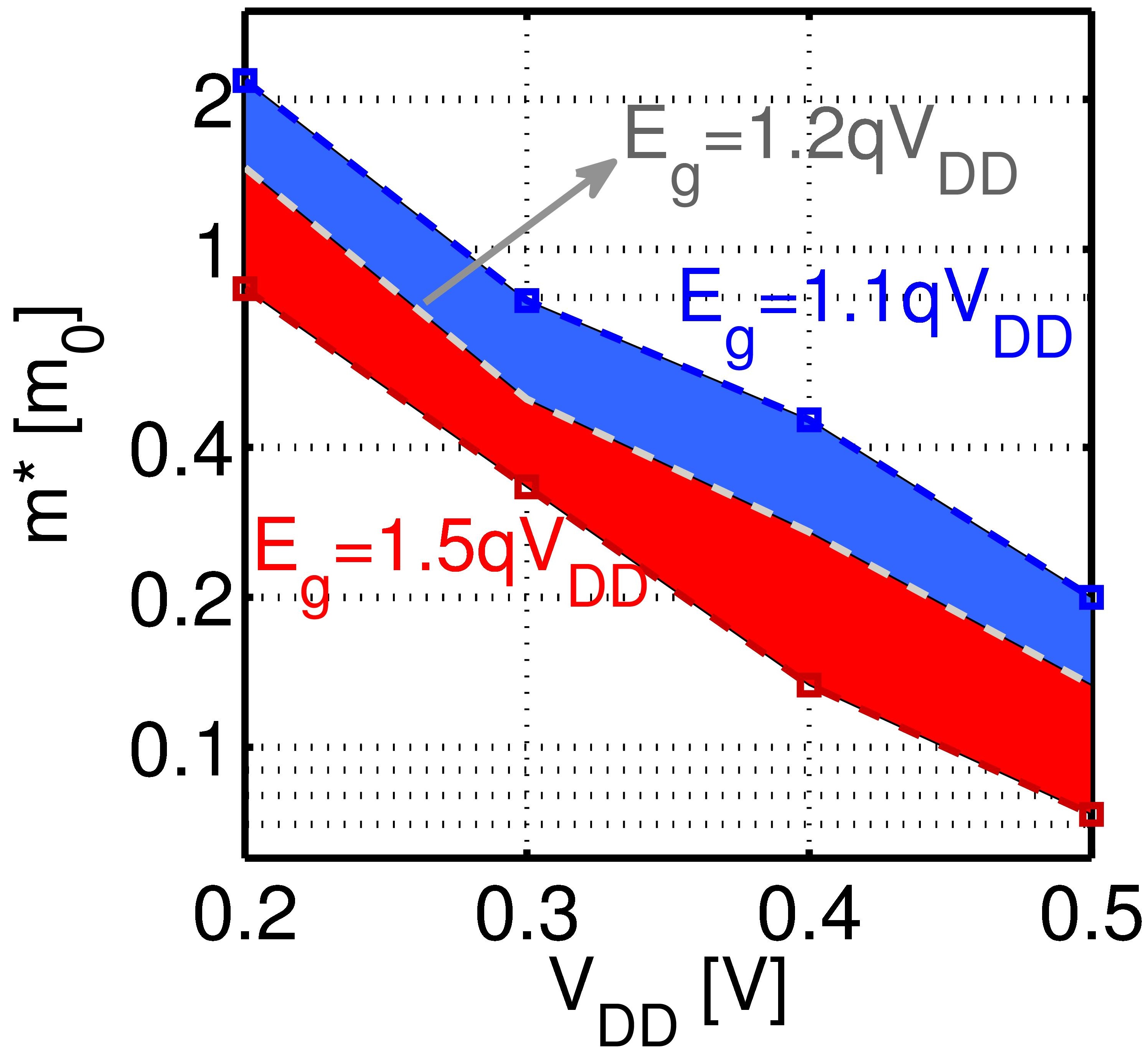} %
               \vspace{-1.5\baselineskip}               
                \caption{}
                \label{fig:laplace}
        \end{subfigure}%
        \begin{subfigure}[b]{0.25\textwidth}
               \includegraphics[width=\textwidth]{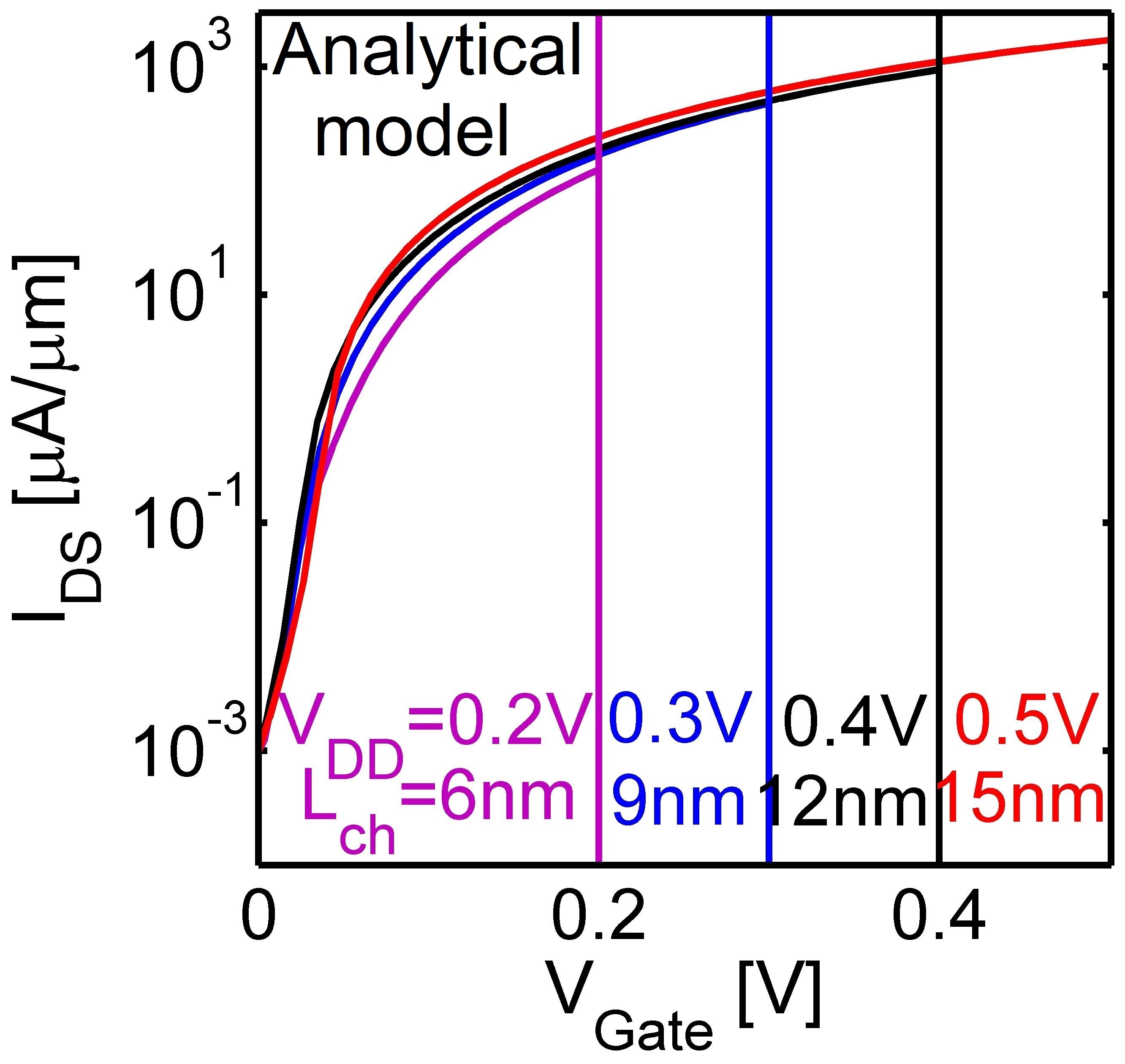}
               \vspace{-1.5\baselineskip}               
                \caption{}
                \label{fig:laplace}
        \end{subfigure}%
        \vspace{-0.5\baselineskip}       
        \caption{a) The optimum effective mass as a function of $V_{DD}$ for TFETs with the scaling rule of $L_{ch}/V_{DD}=30 nm/V$ for $L_{ch}$ from 15nm to 6nm. b) $I_{D}$-$V_{G}$ of NW-TFETs with optimized $E_g$ and $m^*$ from (\ref{eq:opt_VDD}) and (\ref{eq:opt_m}). }\label{fig:Fig5}	
\end{figure}

Fig. \ref{fig:Fig5}a illustrates the favorable design space for $m^*$ as a function of $V_{DD}$ for TFETs with the scaling rule of $L_{ch}/V_{DD}=30 nm/V$. The shaded area in Fig. \ref{fig:Fig5}a shows higher and lower bounds on $m^*$ and $E_g$ of the channel material for a high performance ultra-scaled NW-TFET.
Fig. \ref{fig:Fig5}b shows the transfer characteristics of NW-TFETs with optimized $E_g$ and $m^*$ from equations (\ref{eq:opt_VDD}) and (\ref{eq:opt_m}). I$_{\rm ON}$/I$_{\rm OFF}$ ratio of larger than $10^5$ and $SS$ below 15$mV/decade$ are obtained for all the cases including the 6nm long channel. 
\section{Conclusion}
In summary, the scaling of TFETs pushes the semiconductor industry to look for channel materials with higher $m^*$, similar to ultra-scaled MOSFETs \cite{Sub12}. However, in TFETs channel material should have both $m^*$ and $E_g$ optimized. More accurately, the scaling of high performance NW-TFETs below 10nm requires:
\begin{enumerate}
\item A channel material with \emph{scaled down} band gap
\begin{equation}
\label{eq:opt_VDD}
E_g^{Best}\sim1.2qV_{DD} [eV]
\end{equation}
\item A channel material with \emph{scaled up} effective mass
\begin{equation}
\label{eq:opt_m}
{m^*}^{-1}_{Best}\sim40 V_{DD}^{2.5} [m_0^{-1}]
\end{equation}
\item Higher doping level in the source ($N_S$) than drain ($N_D$).
\begin{equation}
\label{eq:opt_Dop}
N_{S} \gg N_D
\end{equation}
\item A channel material with low dielectric constant ($\epsilon_{ch}$) and a high-k oxide. %
\begin{equation}
\label{eq:opt_Dop}
\epsilon_{ox} \gg \epsilon_{ch}
\end{equation}
\end{enumerate}

\end{document}